\documentclass[pra,twocolumn,showpacs,superscriptaddress,floatfix, reprint]{revtex4}

\usepackage[OT1]{fontenc}
\usepackage[usenames,dvipsnames]{color}
\usepackage[latin1]{inputenc}
\usepackage[english]{babel}
\usepackage{graphicx}
\usepackage{color}
\usepackage{amssymb,amsmath}
\usepackage[Gray,squaren]{SIunits}
\usepackage{xspace}

\usepackage{upgreek}
\usepackage{ulem}
\usepackage{epstopdf}
\usepackage{booktabs}

\begin{document}

\title{Hyperfine measurement of $6\text{P}_{1/2}$ state in $^{87}\text{Rb}$ using double resonance on blue and IR transition}

\author{Elijah Ogaro Nyakang'o}
\address{Department of Physics, Indian Institute of Technology Guwahati, Guwahati, Assam 781039, India}
\author{Dangka Shylla}
\address{Department of Physics, Indian Institute of Technology Guwahati, Guwahati, Assam 781039, India}
\author{Vasant Natarajan}
\address{Department of Physics, Indian Institute of Science, Bangalore, 560012, India}
\author{Kanhaiya Pandey}
\email{kanhaiyapandey@iitg.ac.in}
\address{Department of Physics, Indian Institute of Technology Guwahati, Guwahati, Assam 781039, India}

\date{\today{}}

\begin{abstract}
In this paper, we present the spectroscopy of 6P$_{1/2}$ state in $^{87}$Rb using double resonance technique at $780~\text{nm}$ and $421~\text{nm}$. The double resonance technique is implemented using electromagnetically induced transparency (EIT) and  optical pumping methods. Using these spectroscopy methods, we have measured the hyperfine splitting of 6P$_{1/2}$ state with precision of $<$400~kHz which agrees well with other spectroscopy methods such as electrical discharge and saturated absorption spectroscopy at $421~\text{nm}$.
\end{abstract}


\maketitle

\section{Introduction}

Precise measurements of the hyperfine structure of various lines in an atom provide key information about the properties of the nucleus such as the electric and magnetic moments. Rb is one of the most widely investigated elements in atomic physics for the  spectroscopy both experimentally \cite{YSJ96, MSF05, BAN04, TAC97, PPP13, HMY05} and theoretically \cite{SAS11}. This provide great opportunities to verify different methods of theoretical many-body calculations \cite{CYS18,SSU11}. Hyperfine splitting measurements are good sources of input for studying subjects at the interface of atomic and nuclear physics such as atomic parity violation \cite{SBD18}. Experimentally, hyperfine structures of 5P$_{3/2}$, 5D$_{3/2}$ and 7S$_{1/2}$  have been measured using single-photon transition 5S$_{1/2}$ $\rightarrow$5P$_{3/2}$ at 780~nm \cite{YSJ96, MSF05, BAN04}, and two-photon transitions 5S$_{1/2}$ $\rightarrow$5D$_{3/2}$ at 778~nm \cite{TAC97} and 5S$_{1/2}$ $\rightarrow$ 7S$_{1/2}$ at 760~nm \cite{MSF05, PPP13, HMY05} respectively.

Besides verifying theoretical calculations, the above referred transitions are used as low cost optical frequency standards. For example, the precisely measured transition 5S$_{1/2}$ $\rightarrow$5P$_{3/2}$ at 780~nm is used as an optical reference for measuring unkown transitions \cite{BAN04}. All these transitions fall in IR region; however, the weak and narrow linewidth ($2\pi\times1.27~\text{MHz}$ \cite{SWC04}) transition in the blue region (i.e. at $421~\text{nm}$) has the advantage of high precision for frequency standards \cite{ZZT17,SPH18} and is a promising candidate for metrology. Measuring the hyperfine splitting of $6\text{P}_{1/2}$ adds important input to theoretical calculations \cite{SAS11}. The hyperfine splitting measurement of 6P states has been carried out using saturated absorption \cite{GKG19} for both $6\text{P}_{1/2}$ and $6\text{P}_{3/2}$ states, or fluorescence spectroscopy \cite{NCC19} for $6\text{P}_{3/2}$ state on $5\text{S}_{1/2}\rightarrow6\text{P}_{3/2(1/2)}$ transition and using RF transition with electrical discharge \cite{FDP73}.

The direct detection of absorption of $421~\text{nm}$ on $5\text{S}_{1/2}\rightarrow6\text{P}_{1/2}$ transition requires heating of Rb vapor cell upto $80-100\degree\text{C}$ \cite{GKG19, PBA15} and using a photodiode with blue enhanced sensitivity. The spectroscopy at $421~\text{nm}$ can also be done using double-resonance spectroscopy \cite{BZF98,NYN16,PMH19,SMY88,BPM89} which does not require heating of Rb vapor cell. The double resonance method can be of electromagnetically induced transparency (EIT) type in a V-system \cite{BAV03,DAV05,DAV08,DPW06,MSA99} a technique which is known as coherent control spectroscopy (CCS). We have also added optical pumping technique for the same double resonance spectroscopy. The precise measurement of the hyperfine interval of $6\text{P}_{1/2}$ state in $^{87}\text{Rb}$ is carried out using the two double resonance spectroscopy methods.  
Although the method based upon electrical discharge in reference \cite{FDP73} provides great precision, it is important to measure hyperfine splitting with different methods to avoid systematic shifts in the experiment due to ion-atom and atom-atom collisions. Similarly heating the Rb cell also increases atom-atom collision and can cause collisional/pressure shift \cite{WSY16} which can contribute to systematic shift in the hyperfine measurement.

\section{Measurement Schemes}

\subsection{Coherent Control Scheme}

\begin{figure}
   \begin{center}
      \includegraphics[width=1.0\linewidth]{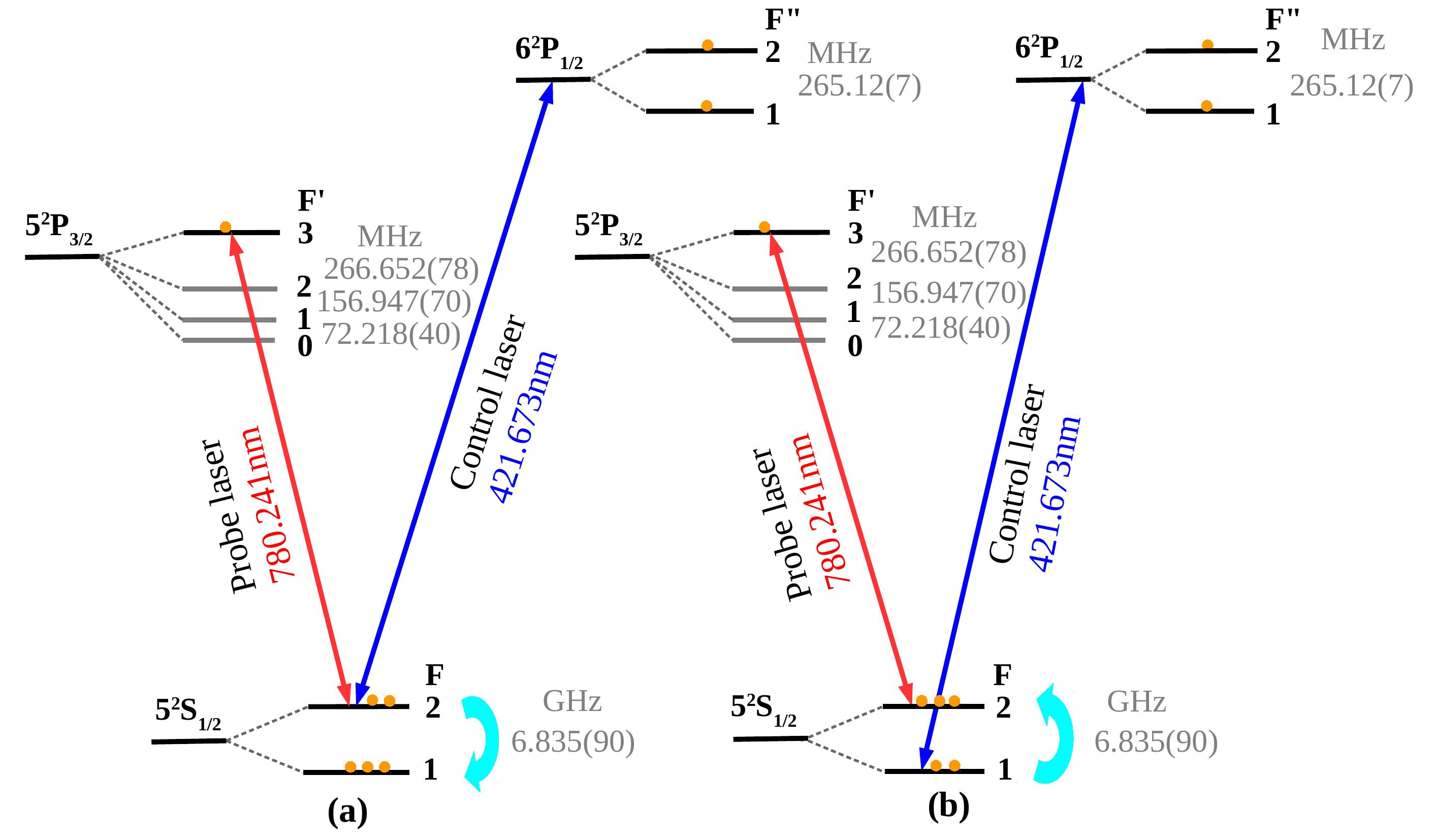}      
      \caption{(Color online). Schematic of a multilevel atomic system interacting with two laser beams in (a) V-type scheme and (b) optical pumping scheme in $^{87}\text{Rb}$.}
      \label{Fig1}
   \end{center}
\end{figure}

The energy level diagram for coherent control scheme is given in Fig. \ref{Fig1}a and the experimental setup is as shown in Fig. \ref{Fig2}. The $780\,\text{nm}$ probe laser is locked to resonance on $5\text{S}_{1/2}(\text{F}=2)\rightarrow5\text{P}_{3/2}(\text{F}'=3)$ cycling transition and its absorption is monitored as the co-propagating $421~\text{nm}$ control laser beam scans $5\text{S}_{1/2}(\text{F}=2)\rightarrow6\text{P}_{1/2}$ transitions. As soon as the $421~\text{nm}$ scanning control laser comes to resonance (i.e. when both laser beams are addressing zero velocity group atoms), absorption of the $780~\text{nm}$ probe laser is reduced giving rise to a Doppler-free dip. There are two reasons for reduction of the 780 nm probe laser absorption. One is due to coherent effect i.e. V system EIT \cite{DAV05,MSA99} and another is optical pumping to other ground hyperfine level, $5\text{S}_{1/2}(\text{F}=1)$ \cite{FBK80,SMH04,HRN09} via $5\text{S}_{1/2}(\text{F}=2)\rightarrow6\text{P}_{1/2}$ excitation and $6\text{P}_{1/2}\rightarrow5\text{S}_{1/2}(\text{F}=1)$ decay channels. The transparency spectrum is shown in Fig. \ref{Fig3}a.

Besides the two hyperfine peaks due to zero velocity group atoms, there are other extra peaks outside the main spectrum. The extra peaks are caused by atoms moving with velocities 208 m/s and 330 m/s respectively. Atoms moving with velocity 208 m/s will see the $780~\text{nm}$ probe laser to be on resonance with $5\text{S}_{1/2}(\text{F}=2)\rightarrow5\text{P}_{3/2}(\text{F}'=2)$ transition. The corresponding two extra peaks are separated by hyperfine interval of 6P$_{1/2}$ and located at 494 MHz from the main peaks respectively. Similarly, atoms moving with velocity 330 m/s will be resonant for $5\text{S}_{1/2}(\text{F}=2)\rightarrow5\text{P}_{3/2}(\text{F}'=1)$ transition and another two fold of extra peaks are located at 783 MHz from the main peaks. The theoretical plot in Fig. \ref{Fig3} is generated using density matrix calculation for seven-level system in Doppler-broadened Rb atomic vapors at room temperature (300~K). Due to non-linearity in the scan of the laser, there is a mismatch between experiment and theory in the position of the extra peaks. The linewidth of the experimental spectrum ranges between $29$ and $31~\text{MHz}$ and the theoretical simulation curve has a linewidth of $26~\text{MHz}$. However, this linewidth is larger than the natural linewidth ($6.065+1.27~\text{MHz}$). This is caused by Doppler mismatch between the $780~\text{nm}$ and $421~\text{nm}$ lasers \cite{UCR13}.

\subsection{Optical Pumping Scheme}

Fig. \ref{Fig1}b is the energy level diagram for optical pumping scheme and the experimental setup is also given in Fig. \ref{Fig2}. The $780\,\text{nm}$ probe laser is locked to resonance on $5\text{S}_{1/2}(\text{F}=2)\rightarrow5\text{P}_{3/2}(\text{F}'=3)$ cycling transition and its absorption is monitored as the co-propagating $421\,\text{nm}$ control laser beam scans around the $6\text{P}_{1/2}$ hyperfine levels on $5\text{S}_{1/2}(\text{F}=1)\rightarrow6\text{P}_{1/2}$ transition instead of $5\text{S}_{1/2}(\text{F}=2)\rightarrow6\text{P}_{1/2}$ transition. The $421\,\text{nm}$ scanning control laser beam, partially transfers population from the lower ground hyperfine level ($5\text{S}_{1/2}(\text{F}=1)$) to the upper ground hyperfine level ($5\text{S}_{1/2}(\text{F}=2)$) via $5\text{S}_{1/2}(\text{F}=1)\rightarrow6\text{P}_{1/2}$ excitation and $6\text{P}_{1/2}\rightarrow5\text{S}_{1/2}(\text{F}=2)$ decay channels. Thus, optical pumping of zero velocity group atoms to the upper ground hyperfine level \cite{FBK80,SMH04,HRN09} and coherence dephasing rate of the ground hyperfine levels \cite{FSM95,MSA14,TSR10} increase absorption of the probe giving rise to Doppler-free peaks. The absorption spectrum is shown in Fig. \ref{Fig3}b. Since all velocity group atoms are optically pumped from $5\text{S}_{1/2}(\text{F}=1)$ to $5\text{S}_{1/2}(\text{F}=2)$ ground hyperfine level, extra peaks are formed outside the main spectrum as explained in the previous section. The linewidth of the experimental spectrum ranges between $29$ and $34~\text{MHz}$ and linewidth for theoretical simulation curve is $23$ and $34~\text{MHz}$.

\section{Experimental Details}

\subsection{Setup and Results}
\begin{figure}
   \begin{center}    
    \includegraphics[width =1.0\linewidth]{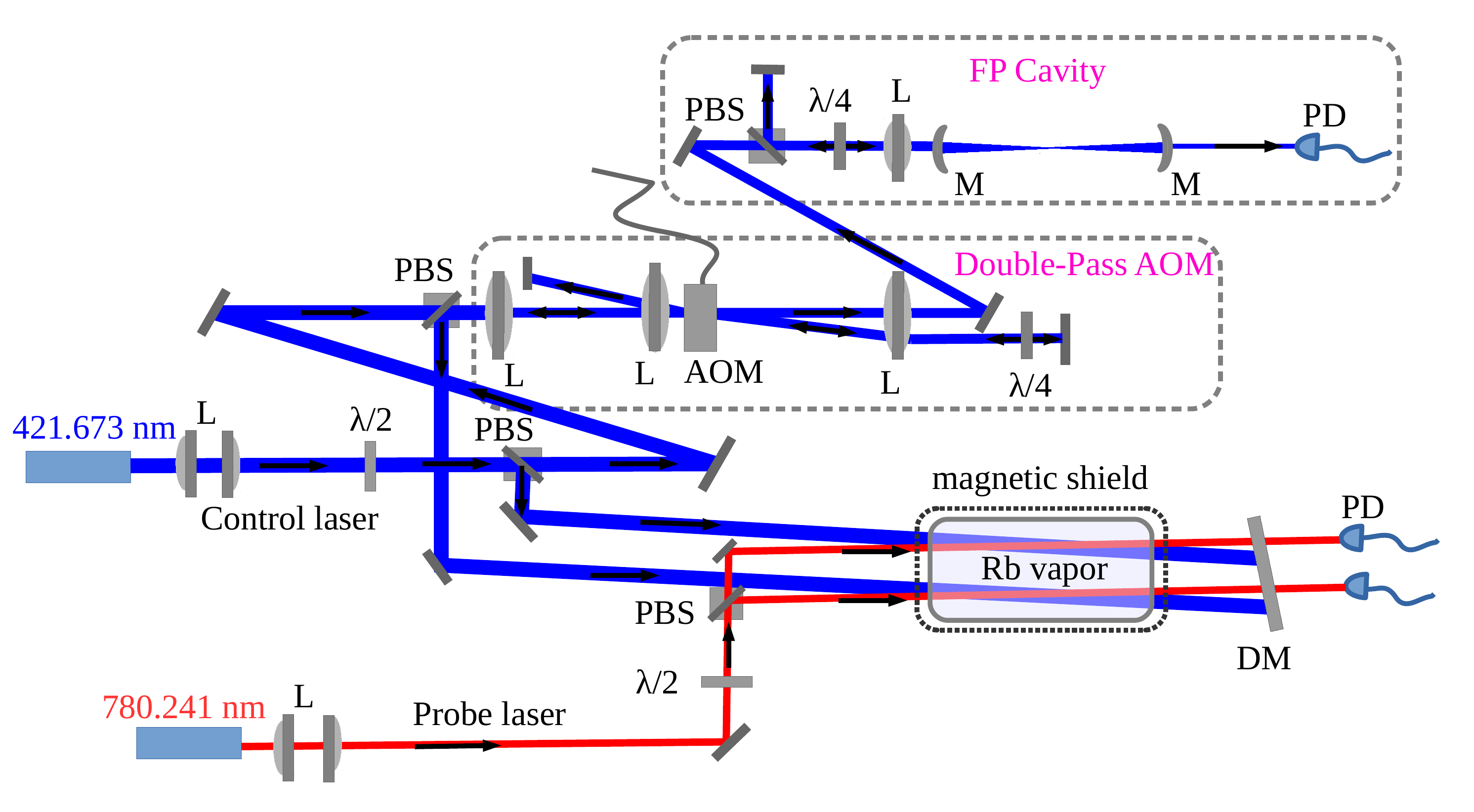}
      \caption{(Color online). Experimental setup for measuring hyperfine structure using coherent control and optical pumping schemes. L: Plano-convex lens; $\lambda/2$: half-wave plate; $\lambda/4$: quarter-wave plate; M: confocal mirror; DM: Dichroic mirror; PBS: polarization beam splitter; PD: photo-diode; AOM: acousto-optic modulator; FP: Fabry Perot cavity.}
      \label{Fig2}
   \end{center}
\end{figure}

The 780 nm beam is generated from (thorlab laser diode L785H1) a home-assembled extended cavity diode laser (ECDL) with typical linewidth of 500~kHz. The error signal for locking the 780 nm laser is generated by frequency modulation using the current of ECDL at 50 kHz. The error is fed to the piezo using a home-made analog PID controller for locking to the particular transition. The 421 nm beam is generated from commercial available ECDL from TOPTICA with model no. DL PRO HP with output power of 70 mW and linewidth of $<$200 kHz. In the experimental setup given in Fig. \ref{Fig2}, the $421\,\text{nm}$ laser beam addressing $6\text{P}_{1/2}$ hyperfine level is divided into two laser beams. The first laser beam is passed directly through the Rb vapor cell and co-propagates with one of the $780\,\text{nm}$ probe laser. The second $421\,\text{nm}$ laser beam is passed through the acousto-optic modulator (AOM) twice and its frequency is shifted to be approximately the hyperfine interval value. The double-pass AOM configuration has the advantage of preserving the direction of propagation of the laser beam as the frequency of AOM is changed \cite{DHL05}. The AOM frequency in our double-pass setup is shifted between $130-136\,\text{MHz}$. The double passed AOM beam, again passes through the same Rb vapor cell where it co-propagate with the second $780~\text{nm}$ probe laser. The two sets of co-propgating 421 nm and 780 nm laser are around 12 mm apart in the same cell. The single-mode operation of the $421~\text{nm}$ laser is monitored using Confocal Fabry Perot Interferometer with free spectral range of 150~MHz. The beam diameter of the $780~\text{nm}$ probe laser is $2\times 3~\text{mm}$ with measured power of $42~\mu\text{W}$ (or intensity, $\text{I}=1.78~\text{mW}/\text{cm}^2$ and corresponding Rabi frequency of $2\pi\times4.27~\text{MHz}$). The beam diameter of the $421\,\text{nm}$ control laser is $3\times 4~\text{mm}$ with measured power of $0.945~\text{mW}$ and calculated intensity, $\text{I}=20.05~\text{mW}/\text{cm}^2$. The intensity corresponds to Rabi frequency of $2\pi\times1.17~\text{MHz}$ using the dipole moment in reference \cite{SWC04}

\begin{figure}
   \begin{center}    
    \includegraphics[width =1.0\linewidth]{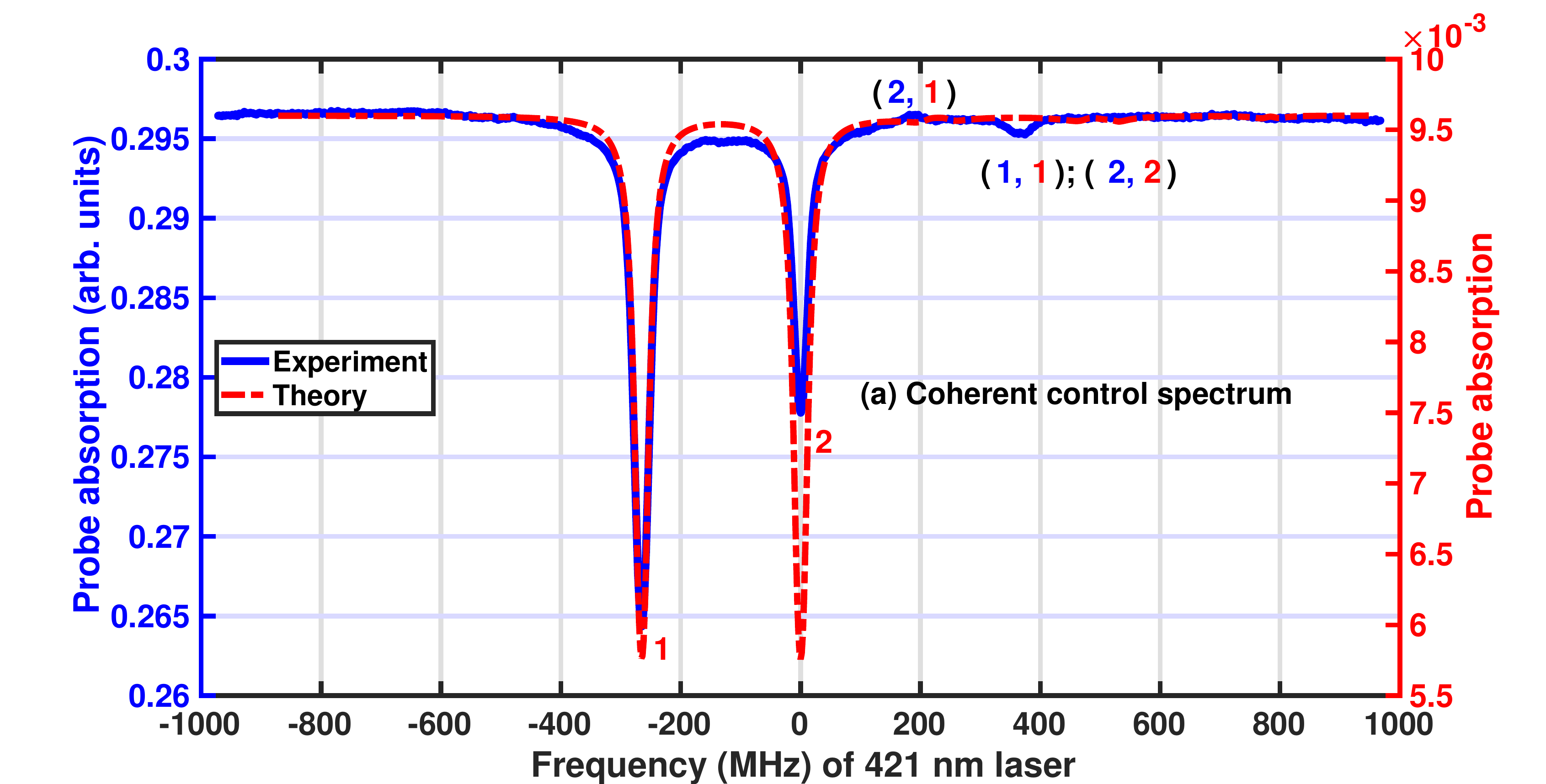}
    \includegraphics[width =1.0\linewidth]{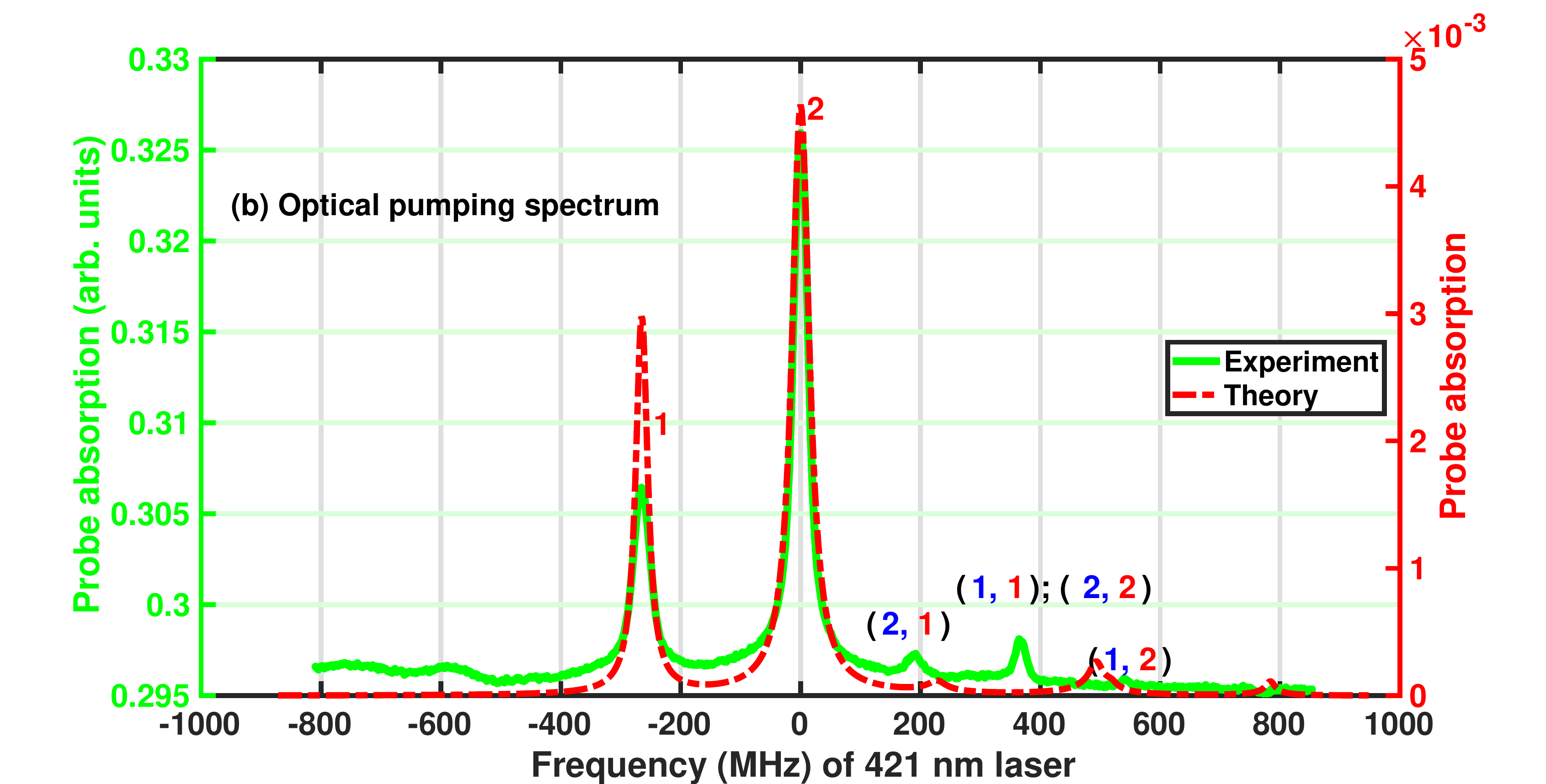}
      \caption{(Color online). Theoretical and experimental spectrum of $6\text{P}_{1/2}$. Extra peaks are caused by atoms moving with velocity $208~\text{m/s}$ and $330~\text{m/s}$ which brings 780 nm and 421 nm lasers to resonance on $5\text{S}_{1/2}(\text{F}=2)\leftrightarrow5\text{P}_{3/2}(\text{F}'=1,2)$ (blue color) and $5\text{S}_{1/2}(\text{F}=1(2))\leftrightarrow6\text{P}_{1/2}(\text{F}''=1,2)$ transition (red color).}
      \label{Fig3}
   \end{center}
\end{figure}

\begin{figure}
   \begin{center}
      \includegraphics[width=1.0\linewidth]{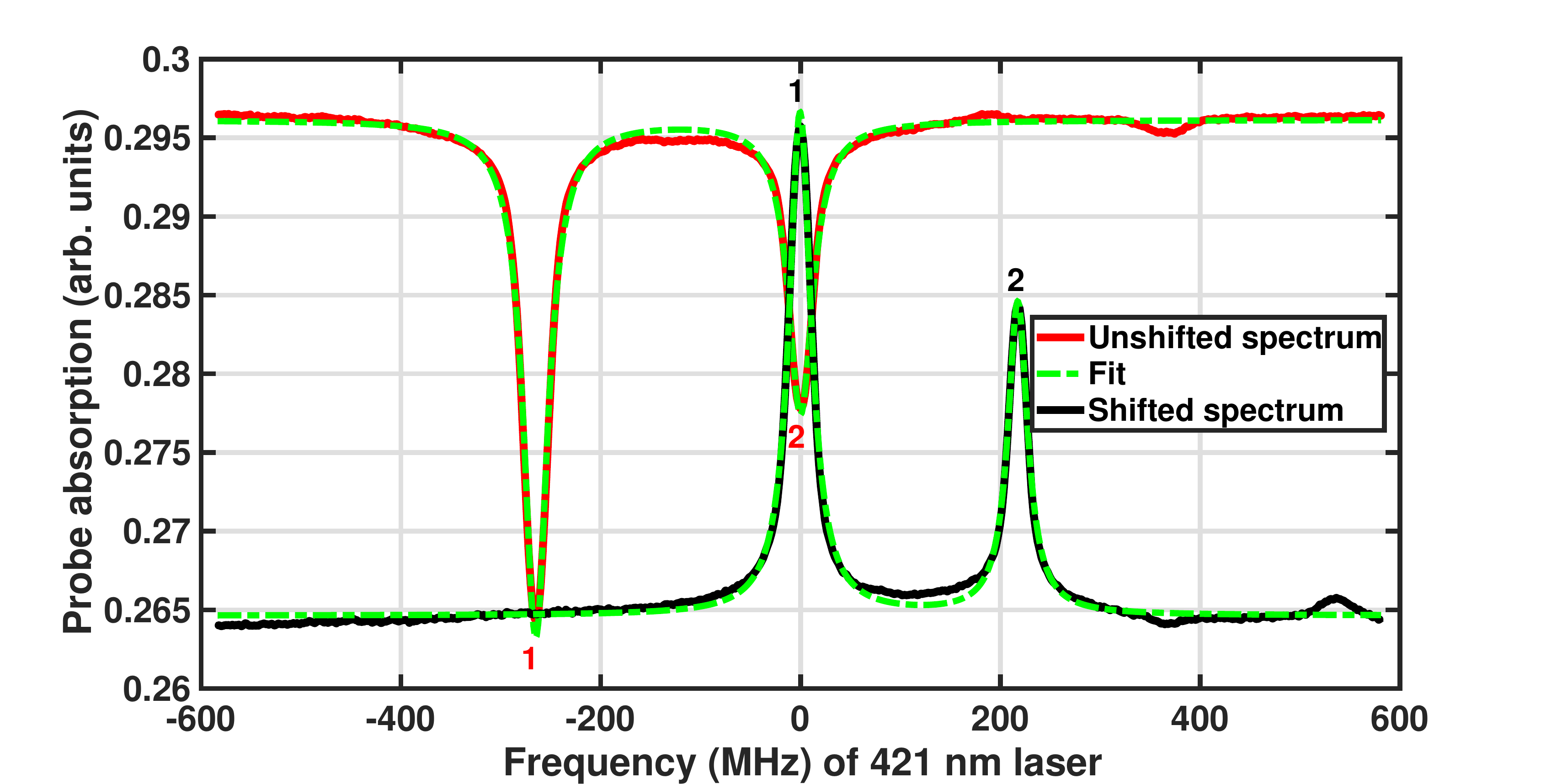}      
      \caption{(Color online). Spectrum of shifted (black color) and unshifted (red color) beams fitted with a Lorentzian line profile (dashed green color) to obtain frequency difference ($\Delta_{\textrm{diff}}$) between the matched peaks.}
      \label{Fig4}
   \end{center}
\end{figure}

The spectrum of $5\text{S}_{1/2}(\text{F}=2)\rightarrow6\text{P}_{1/2}$ or $5\text{S}_{1/2}(\text{F}=1)\rightarrow6\text{P}_{1/2}$ weak transition driven by $421~\text{nm}$ laser shown in Fig. \ref{Fig3}a and \ref{Fig3}b respectively, is recorded using a picoscope through the changes in the absorption spectrum of $780~\text{nm}$ probe laser driving $5\text{S}_{1/2}(\text{F}=2)\rightarrow5\text{P}_{3/2}(\text{F}'=3)$ strong transition. The red and black traces of experimental spectrum in Fig. \ref{Fig4} corresponds to unshifted and shifted AOM beams respectively. One of the traces is deliberately inverted to see the matching of the two hyperfine peaks for the shifted and unshifted spectrum. The matching of the peaks is a measure of shifting the frequency of the laser beam by exactly the hyperfine interval. The frequency difference ($\Delta_{\textrm{diff}}$) between the two peaks being matched is obtained by fitting the peaks with a Lorentzian line profile (see Fig. \ref{Fig4}) and finding the difference in the peaks location. Fig. \ref{Fig5} shows a plot of frequency shift (2$\times$ AOM frequency) vs the frequency difference between the two peaks ($\Delta_{\textrm{diff}}$). The hyperfine interval is obtained using a linear fit on the plot of frequency shift vs $\Delta_{\textrm{diff}}$. The frequency shift corresponding to zero frequency difference ($\Delta_{\textrm{diff}}=0$) in the linear fit is the hyperfine interval ($\mathcal{V}_\text{hfs}$). This method removes the error due to scan non-linearity and hence improves the precision of measurement. From the linear fit the value of $\mathcal{V}_\text{hfs}=265.134\pm0.047~\text{MHz}$ in the case of the coherent control scheme and $\mathcal{V}_\text{hfs}=265.196\pm0.034~\text{MHz}$ for the optical pumping scheme. 

\subsection{Errors}

\subsubsection{Systematic Errors}

The main source of the systematic errors is the light shift and stray magnetic field through Zeeman shift. The systematic error arising due to stray magnetic field is minimized using a $\mu$-metal magnetic shield around the Rb cell. The residual fields is below 1 mG which corresponds to errors less than 1 kHz. The light shift error is due to presence of the hyperfine levels and the lasers driving simultaneously many levels off resonance causing the light shift of the levels driven resonantly. The locked probe laser $5\text{S}_{1/2}(\text{F}=2)\rightarrow5\text{P}_{3/2}(\text{F}'=3)$ cycling transition, also drives $5\text{S}_{1/2}(\text{F}=2)\rightarrow5\text{P}_{3/2}(\text{F}'=2(1))$ transitions off resonance causing the light shift to the ground state $5\text{S}_{1/2}(\text{F}=2)$ upward and excited state $5\text{P}_{3/2}(\text{F}=3)$ downwards. However, this shift does not cause any error for hyperfine interval because it will cause equal shift in the resonance for both the hyperfine levels of $6\text{P}_{1/2}$. The scanning control laser is the source of systematic error in the measurement of hyperfine interval. This is because, when it is resonant to $5\text{S}_{1/2}(\text{F}=2)\rightarrow6\text{P}_{1/2}(\text{F}''=1)$, it also driving the $5\text{S}_{1/2}(\text{F}=2)\rightarrow6\text{P}_{1/2}(\text{F}''=2)$ off resonance (negative detuning equal to hyperfine interval, $\mathcal{V}_\text{hfs}$) causing the ground state $5\text{S}_{1/2}(\text{F}=2)$ to shift downwards by $\Omega^2/4\mathcal{V}_\text{hfs}$. This effect causes resonant frequency for $5\text{S}_{1/2}(\text{F}=2)\rightarrow6\text{P}_{1/2}(\text{F}''=1)$ to be shifted by $+\Omega^2/4\mathcal{V}_\text{hfs}$. Similarly, when the control laser is at resonance on $5\text{S}_{1/2}(\text{F}=2)\rightarrow6\text{P}_{1/2}(\text{F}''=2)$ transition, it is also driving the $5\text{S}_{1/2}(\text{F}=2)\rightarrow6\text{P}_{1/2}(\text{F}''=1)$ transition off resonance (positive detuning equal to hyperfine interval) causing the ground state $5\text{S}_{1/2}(\text{F}=2)$ to shift upwards by $\Omega^2/4\mathcal{V}_\text{hfs}$. This causes resonant frequency for $5\text{S}_{1/2}(\text{F}=2)\rightarrow6\text{P}_{1/2}(\text{F}''=2)$ transition to be shifted by $-\Omega^2/4\mathcal{V}_\text{hfs}$. The overall light shift error calculated using the laser intensities in the previous section is $13~\text{kHz}$ and $6~\text{kHz}$ for the coherent control scheme and the optical pumping scheme respectively.

\subsubsection{Statistical Error}

\begin{figure}
   \begin{center}
      \includegraphics[width=1.0\linewidth]{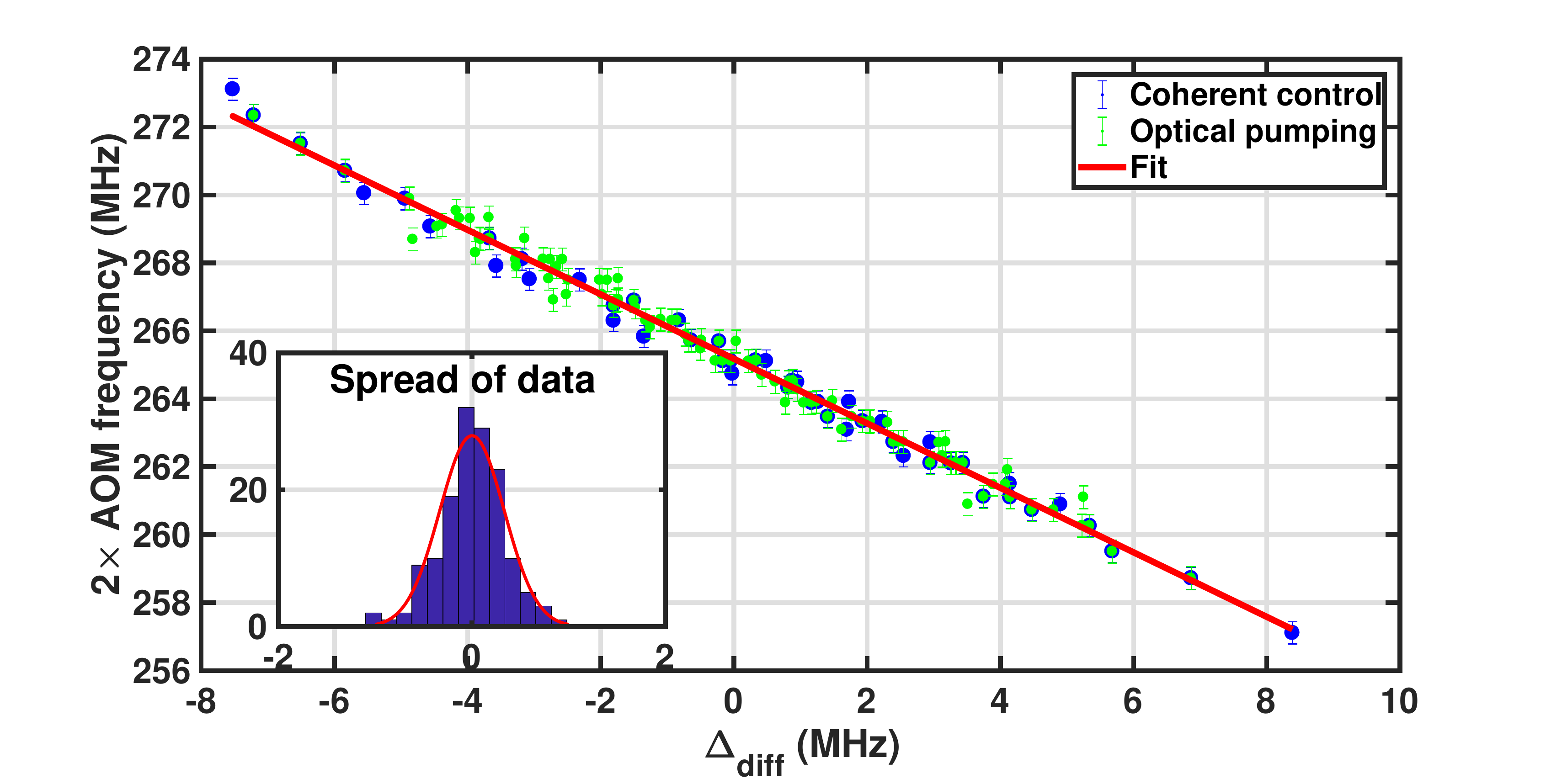}      
      \caption{(Color online). A plot of frequency shift (2$\times$ AOM frequency) vs frequency difference ($\Delta_{\textrm{diff}}$) for the two schemes. The inset shows the spread of data from the mean hyperfine interval.}
      \label{Fig5}
   \end{center}
\end{figure}

The above systematic error is much smaller than the statistical error in the experiment. The non-linear scan of the laser is the main cause the statistical error. This error is minimized by shifting AOM frequency within a small range of frequencies around the neighboring hyperfine level. To quantify the statistical error, two traces (shifted and unshifted spectrum) are recorded on two input channels of the picoscope with averaging of 20. Three such samples are taken for each AOM frequencies and the spread of the data ($\Delta_{\textrm{diff}}$) is shown by the histogram in the inset of Fig. \ref{Fig5}. The spread of the data gives the statistical error in the experiment and is extracted from the histogram using a Gaussian fit. The extracted statistical error is $0.326$ MHz for the coherent control scheme and $0.337$ MHz for the optical pumping scheme. 

\begin{table}[ht]
\centering
\caption{Hyperfine interval ($\mathcal{V}_\text{hfs}$) and magnetic dipole constant A for $6\text{P}_{1/2}$ state in $^{87}\text{Rb}$. The number indicated in normal bracket is the statistical plus fitting error and in curly bracket is systematic error.}
\scriptsize{}
\begin{tabular}[t]{lccc}
\toprule
&$\mathcal{V}_\text{hfs}\,$(MHz)&A (MHz)&Reference\\
\midrule
Coherent control&265.134(373)\{14\}&132.567(200)&This work\\
Optical pumping&265.196(371)\{7\}&132.598(200)&This work\\
&265.150(460)&132.83(500)&\cite{GKG19}\\
&265&&\cite{GGR77}\\
&&132.56(3)&\cite{FDP73,AIV77}\\
\bottomrule
\end{tabular}
\label{tab:1}
\end{table}

In summary, the statistical error is dominating over systematic errors (light shift and stray magnetic field errors) and fitting error. The total error is $0.387~\text{MHz}$ and $0.378~\text{MHz}$ for the coherent control scheme and the optical pumping scheme respectively. Hence the hyperfine interval in the case of coherent control scheme is $\mathcal{V}_\text{hfs}=265.134(373)\{14\}$ MHz and optical pumping scheme is $\mathcal{V}_\text{hfs}=265.196(371)\{7\}$ MHz. The measured hyperfine interval is related to the magnetic dipole hyperfine constant, $\text{A}=\mathcal{V}_\text{hfs}(\text{F}\rightarrow\text{F}-1)/\text{F}$. The values of A are $132.567(200)$ MHz and $132.598(200)$ MHz for the two schemes respectively. A comparison of hyperfine interval ($\mathcal{V}_\text{hfs}$) and magnetic dipole constant A with the earlier works is given in Tab. \ref{tab:1}.

\section{Conclusions}

We have presented two experimental schemes for precision measurement of hyperfine interval of $6\text{P}_{1/2}$ state of $^{87}\text{Rb}$, namely coherent control and optical pumping schemes using double resonance at 780~nm and 421~nm. Using an AOM, we have taken care of the scan non-linearity which is the dominant source of error in the experiment. The measured hyperfine interval is consistent with two other techniques namely saturated absorption and electrical discharge within the precision of our measurement.

\section*{Acknowledgement}
E.O.N. would like to acknowledge Indian Council for Cultural Relations (ICCR) for the PhD scholarship. K.P. would like to acknowledge the funding from SERB of grant No. ECR/2017/000781.

\bibliography{CCSpaper}

\end{document}